*Observation of Spontaneous Helielectric Nematic Fluids: Electric Analogy to Helimagnets*

Xiuhu Zhao[1]†, Junchen Zhou[1]†, Hiroya Nishikawa[2]†, Jinxing Li[1], Junichi Kougo[1,3], Zhe Wan[1], Mingjun Huang[1,3]*, Satoshi Aya[1,3]*

† X. Z., J. Z., H.N. contributed equally to this work.

[1] *Department: South China Advanced Institute for Soft Matter Science and Technology (AISMST), School of Molecular Science and Engineering. South China University of Technology, China.*

[2] *Physicochemical Soft Matter Research Team, RIKEN Center for Emergent Matter Science (CEMS), Japan.*

[3] *Guangdong Provincial Key Laboratory of Functional and Intelligent Hybrid Materials and Devices, South China University of Technology, Guangzhou 510640, China.*

*Abstract*

About a century ago, Born proposed a possible matter of state, ferroelectric fluid, might exist if the dipole moment is strong enough. The experimental realisation of such states needs magnifying molecular polar nature to macroscopic scales in liquids. Here, we report on the discovery of a novel chiral liquid matter state, dubbed chiral ferronematic, stabilized by the local ferroelectric ordering coupled to the chiral helicity. It carries the polar vector rotating helically, corresponding to a helieletric structure, analogous to the magnetic counterpart of helimagnet. The state can be retained down to room-temperature and demonstrates gigantic dielectric and nonlinear optical responses. The novel matter state opens a new chapter for exploring the material space of the diverse ferroelectric liquids.

*Introduction*

Among nature, a new matter state usually arises as a result of unexpected combinations of hierarchical orderings of matters. Such discoveries throughout human history have greatly enhanced our understanding of nature and provided us with numerous opportunities to leverage the underlying mechanism for novel material design. Helicity is one of the most essential nature of matter state for organising superstructures in soft matters, spanning on many length scales from the atomic to the macroscopic biological levels. When constructed from building blocks that have inherent polarity, three hierarchical orderings could coexist in a helical structure: 1) the head-to-tail or polar symmetry of each building block; 2) the orientational order of a swarm of building blocks; 3) the emergent helicity (Fig. 1B and 1C); While a simultaneous realization of these three orderings could lead to extraordinary material properties, such highly hierarchical structures are often challenging to achieve in man-made systems. Probably the most familiar example is the chiral magnet or helimagnet (Fig. 1B), and skyrmion (Fig. 1C) in quantum systems, where magnetic spins form 2D or 3D spiral structures (*1-4*)]. The polar magnetic helical structures are considered mainly to originate from either the breaking of the space-inversion symmetry of crystal structures [5] or the magnetic frustration [3,6,7]. Their strong magnetism-chirality coupling triggers enormous interests in condensed-matter physics, leading to many unique functionalities such as extraordinary Hall effect [8], ferroelectric-polarization-induction [9] and high-density information storage []10,11. From the mirror relationship between the magnetism and electricity, we anticipate the incidence of a possible electric version of the helimagnets would intrigue emergent phenomena in condensed-matter physics, as well as additional dimensions to create novel ferroelectric devices. However, the diverse magnetic topological states rarely show up in electric systems, except a few recent breakthroughs, e.g. the observation of the electric skyrmions, polar vortices and merons in metal-organic crystals [12-14]. The special electric states at nanoscale represent extraordinary properties such as local negative dielectric permittivity [15] and strain–polarization coupling [16,17]. Moreover, nearly all the aforementioned helimagnet or electric-analog systems are based on elaborately fabricated inorganics. It is expected that the revolutionary realization of these topologies in soft matter system would bring the advantages of flexibility, simple preparation, large-area film formation and ease integration into electric devices. Here we report on a discovery of such a state in fluidic liquid crystals (LCs), dubbed helielectric nematic, where the spontaneous polar nematic ordering is coupled to the chiral orientational helicity. In contrast to the traditional nanoscopic helimagnetic or helielectric state, a wide tunability of the periodic distance of the fluidic structure in the range of micrometres to near-ultraviolet wavelength region is achieved. Besides, the ability of the switching between the polar and nonpolar helical matter states allows us to systematically study at what conditions and how the particular state is stabilised. As gifts of the chirality-polarity interaction, the matter state uniquely expresses giant dielectric and nonlinear second-harmonic generation optical response.

*Creation and observation of helielectric fluids with nematic order* – Fluids with the traditional nematic order exhibit the orientational cooperativity but lose the intermolecular positional regularity. The average direction of the local molecular orientation is characterised by the director, n, with the head-to-tail invariant nature. Though a single anisotropic entity (usually a molecule) has a preferred directional dipole, the equivalent number of upwards and downwards dipole vectors in the nematic state results in the zero spontaneous polarisation (Fig. 1D). Introducing strong molecular dipole and geometrical asymmetry into the entity is an effective root for breaking this symmetry, realising a fluidic ferroelectric nematic ($N_F$) state exhibiting the local ferroelectric ordering with orientated polarity [18-20]. The head-to-tail invariance breaks down, so ***n≠-n*** (Fig. 1A). Considering the helimagnet is a helically-modulated ferromagnetic state (Fig. 1B), we conceive the idea to incorporate a chiral generator to induce helicity to the $N_F$ state, leading to the creation of the corresponding helielectric state. To test this possibility, we used a ferroelectric material (RM734) [19], and synthesized S1, S2, E1 and E2 as chiral generators (Figs. 1G and S2). The four chiral generators were designed to incorporate a series of chiral groups at various positions on the molecular scaffold of RM734, giving us the chance to investigate the geometrical effect of the chiral generator. In addition, we also examined a commercial chiral generator S811 as a reference molecule that is structurally distinct from RM734 (Fig. S9). Taken together, these five chiral generators allowed us to systematically investigate how varying structures would affect their helical twisting ability.

To our delight, while the addition of S811 results in noneminent polarity and a strong preference to form crystals, the introduction of S1, S2, E1, E2 indeed facilitates the formation of a novel helielectric state that could exist across a wide temperature window and even at room temperature. Figure 1H demonstrates how the matter state varies as a function of temperature and the weight percentage of the chiral generators (S1 and S2; Materials and Methods). Interestingly, the state diagrams are independent of the molecular structure of the chiral generator and so the helical twisting power (Supplementary Discussions 1). The high weight doping of chiral generators, at least up to 70%, into RM734 still generates the helielectric state, suggesting that the macroscopic polarity is achieved by long-range polar interactions. They show a general trend of the matter phase sequence: isotropic liquid → nonpolar helical nematic state (N*) → helielectric nematic state (HN*) → crystal. The temperature range of the N* phase is linearly reduced with increasing the weight of the chiral generators. It is worth noting that the crystal phase is suppressed when the content of the chiral generators exceeds c.a. 30 percent in weight, resulting in the formation of HN* at room temperature.

Polarized light microscropy (PLM) enabled us to directly visualize the transition from N* state to HN*. Figure 2A shows the evolution of PLM images for RM734/S1=95/5 near the N*-HN* transition in a wedged LC slab, where the so-called Grandjean-Cano lines appear, which represent the disclinations separating aligned helical structures with differed number of the helix [21]. The periodicity $L$ of the Grandjean-Cano lines, which is proportional to the helical pitch $p$ of the nonpolar helical structure in a $\alpha$-degree wedged LC slab following the relationship $p = 2L\tan\alpha$ [21], doubles from the N* to HN* state. Since both the reflection wavelength measurements (Figs. 2E, S7B, S8B and S9B) and microscopy observations (Fig. 3D, *vide infra*) confirmed that the real helical pitch remained almost the same from the N* to HN* state, the observed change of $L$ suggested that the relationship $p = 2L\tan\alpha$ does not apply to the HN* state. In the N* state, the relationship $p = 2L\tan\alpha$ originates from the insertion of a half helical pitch from domain to domain (Fig. 2F, Top) to form the Grandjean-Cano lines. In contrast, the same insertion mode would lead to surface defects in HN* materials given the mismatch between the neighbouring vectorised field of the polarisation. We hypothesized that HN* materials form the Grandjean-Cano lines by inserting a full helical pitch in neighbours to avoid free-energy penalty caused by additional surface defects. Figures 2F,G demonstrate the simulated topological structures in both the N* and HN* states confined in the wedge LC slab (Materials and Methods). Indeed, the HN* state carries disclination lines with a winding number of -1/2. Inserting a full-pitch helielectric structure step-by-step upon increasing the slab thickness has lower free-energy than the counterpart of inserting a half-pitch helielectric structure, leading to the reduction of the density of the disclination lines to half compared to that in the N* state. Therefore, we conclude that the equation describing the relationship between the helical pitch and periodicity of Grandjean-Cano lines transmute into $p = L\tan\alpha$ in the HN* state.

To investigate the emergent polar property of the materials, we conducted the second harmonic generation (SHG) measurement. It probes the existence of the inversion center in the system, characterizing the static polar ordering. Having shown that the helielectric state can be generated at room temperature with a wide tunable range of mixing ratios (Fig. 2B-E), we explored how the mixing ratio (*i.e.* helical pitch) and temperature would affect the SHG. Figure 3A demonstrates the temperature dependencies of the strength of the SH signals for the HN* state with different concentration of chiral generators, S1. For all the mixture materials irrespective of the helical pitch, the SH signal shows up slightly above the N*-HN* transition and increases with decreasing the temperature. The SH signal in the HN* state is steady or slightly increasing for most of the mixture materials, suggesting the emergent polarity is stable with temperature in the HN* state. At lower concentration of S1, the HN* state shows strong SH signal (e.g. maximumly 83, 32 and 0.4 times of SH signal of a Y-cut quartz plate). Increasing the concentration of S1 results in a rapid decrease of the SH signal. This is attributed to the decrease of both the helical pitch (Fig. 3B) and the inherent polarity of the systems by increasing the ratio of the less-polar chiral generators (Fig. 4A-C, Fig. S3, *vide infra*). Especially, as the helical pitch becomes shorter compared to the wavelength of the fundamental ray, the net polarity for the fundamental ray decreases, causing the SH signal weakens. Surprisingly, a clear SHG enhancement up to c.a. 500 times is observed when the concentration of S1 locates in a limited window of 58-62 wt% (inset on right top in Fig. 3A and Fig. 3B). In this range, the helical pitch is about 510-550 nm, comparable to the SH wavelength of 532 nm. The SHG light is circularly polarised following the handness of the helielectric structure. The absolute value of the so-called dissymmetry factor *g*-value, defined as $g = 2(\eta_L - \eta_R)/(\eta_L + \eta_R)$ expressing the purity of the circular polarisation, is c.a. 1.7 for all the tested material, indicating the excellent selectivity of the handedness of the spontaneous SH light generation.

The SH signal is the strongest when the incident polarization of the fundamental ray coincides to the nematic director **n** (inset on left in Fig. 3A). Therefore, the polarity vector is parallel to local **n**. This well-correspondence between the molecular orientation, i.e. the polarity vector, and the magnitude of the SH signal makes the SHG measurement to be a robust method to determine the polarity structure of the HN* state. To visualize the polar helicity, we conducted SHG interference (SHG-I) microscopy that is sensitive to the orientation of the polarity and the sole technique being able to probe the polar properties in bulk fluidic materials [22-25]. Figures 3C-F demonstrate the simultaneous observation of SHG and SHG-I microscopies. The helical axis is aligned parallel to the LC slab surface (Fig. 3C). This condition allows us to directly determine how the polarity vector modulates along the helical axis. The SHG microscopy image reveals that the periodicity of the fine bright stripes is consistent to the width, ~1.1 μm, of the finger texture of both the N* and HN* state. The periodicity is the half of the helical pitch, 2.2 μm, of the HN* state. The simultaneous SHG and SHG-I microscopies in an image clarifies the adjacent bright stripes have a mutual phase difference of π, *i.e.* the alternation of dark and bright in interfered SH signal intensity (Fig. 3E). Figure 3F demonstrates the intensity plots of the cross sections for SHG (line a-a') and SHG-I (line b-b') microscopies. The spatial distributions of both the signals are well fitted, where the two beams are assumed to be impinged onto the CCD as gaussian spots. It confirms the helicity of the polarity vector field as calculated in the top of Fig. 3E. Such a helielectric polar vector field can respond to an electric field. We confirmed it by SHG measurement under an in-plane electric field (IPE-field, refer to Fig. 3G). The polarization of the incident fundamental beam is set to be parallel to the direction of the IPE-field. Upon increasing the electric field, the increase of SH signal was confirmed, indicating the reorientation of the polar vector towards the direction of the IPE-field (Fig. 3H). Over about $9V_{DC}$, electrohydrodynamic convection flow occurs. These results confirm the proposed helielectric state come into existence.

To explain the emergence of the polarity through the N*-HN* phase transition, we learn from the computational models of magnetic systems and describe a Landau–de Gennes type of free-energy density near the phase transition as follows by assuming the polarity **P** as the primary order parameter for the transition:

$$f = a(T - T_c)\mathbf{P}^2 + b\mathbf{P}^4 + 1/2 K_{22}\big(\mathbf{n} \cdot (\nabla \times \mathbf{n})\big)^2 - 1/2 K'_{22}\mathbf{n} \cdot (\nabla \times \mathbf{n}) + J(\nabla \mathbf{P})^2 + D\mathbf{P} \cdot (\nabla \times \mathbf{P}),$$

where $T_c$, P, $K_{22}$ and $K'_{22}$, $q_0$ and **n** denote the transition temperature, polarisation strength, twist elastic coefficient, wave number of the helical pitch and the director vector. The first three terms represent the Landau free-energy, second from the last term the free-energies from the spatial variation of the polarisation field, and the last term corresponds to the electric counterpart of the Heisenberg and DM interactions that represents the tendency of the polarity winding. $(\nabla \mathbf{P})^2$ means $(\partial_i P_j)^2$ in Einstein notation. The polarisation vector is parallel to the director, so $\mathbf{P} = P_0 \mathbf{n}$ as confirmed by SHG measurement. We consider the helical axis oriented along the *z*-direction and use the ansatz for the director field,

$$\mathbf{n} = (\cos q, \sin q, 0).$$

The difference of the free-energy between the nonpolar-polar helical structures is $\Delta f = f(P) - f(0)$. By putting the director field into the free-energy density, $\Delta f$ reads

$\Delta f = a(T - T_c)P^2 + bP^4 + P^2 q(Jq - D)$.

By minimising the free-energy ($\partial \Delta f / \partial P = 0$, $\partial^2 \Delta f / \partial P^2 > 0$), we obtain the solutions for P as $|P| = 0, \sqrt{Dq - Jq^2 - a(T - T_c)}/\sqrt{2b}$.

When $T < T_c + (Dq - Jq^2)/a$, non-zero spontaneous polarization $P$ appears and the free-energy for the HN* state is always lower than that in the N* state by $(Dq - Jq^2 - a(T - T_c))^2/4b$. This suggests the incidence of the HN* would be a general phenomenon if the existence of the polarity is allowed in fluidic systems.

*Liquid capacitor with extraordinary dielectric properties* – In contrast to solid-based dielectrics, the helielectric materials can be easily filled between flexible substrates and deformed, possibly serving as a promising flatform for liquid capacitors. Considering the huge polarity of the HN* materials reflected in the SHG response, it is of great interest to see how strong the dielectric response that comes from the dynamic polarity switching is, and in what range the capacitance can be controlled as electronic elements. We investigated the temperature dependence of the dielectric response for materials with different concentration of the chiral generator S1 (Fig. 4A-C). Surprisingly, the giant dielectric constant in the range of 1000-10000 at 1000-500 Hz is observed for the mixtures with S1 below 20 wt% (Fig. 4B). Similar to the SHG measurement results, upon increasing the doping ratio of the chiral generator, the maximum dielectric constant decreases (Fig. 4B). Thanks to the huge dielectric constant, the capacitance of the HN* state is in the range of nano-farad. For example, the effective capacitance of a HN* capacitor at RM734/S1=90/10 contacting with 1cm × 1cm electrodes is as huge as 20 nF when the film thickness is c.a. 100 μm, and is dramatically up-tuned to c.a. 200 nF by decreasing the film thickness down to c.a. 5 μm (inset of Fig. 4B). The dielectric constant for the mixtures increases when approaching the N*-HN* transition and then decreases (Fig. 4C), *i.e.* a bell-shape temperature variation. Worth noting, the polarity never weakens in the HN* by lowering temperature, which is manifested in the steady increase of the SH signal (Fig. 3A). Therefore, the bell-shape temperature variation of dielectricity would be interpreted as below. First, the nucleation of and growth of microscopic polar domains embedded in the nonpolar background proceed with lowering temperature, giving rise to the increase of both the SHG and dielectric constant. Under this circumstance, the average polar vector in each polar domain is small and can be switched by the probe electric field (typically 10-100 mV). At lower temperatures, each polar domain becomes too big to respond to the probe electric field during the dielectric measurement. This is consistent with the scenario of the relaxor-like behaviours [26,27].

In summary, we report the first real-space observation of polar helielectric state in nematic fluids. The realization of such a state with the homochiral helical arrangement of the polarity vector is made possible through the coupling between the strong local polarity and chirality. This unique electric structure in fluids resembles the helimagnets in the context of magnetic systems, but distinct in their high tunability of the helielectric pitch in a wide range from hundred nanometers up to several tens of micrometer. The stabilization gifts the giant dielectric susceptibility and amplified nonlinear optical effects. The discovery opens the door for exploring physics of novel polar topological structures previously inaccessible in fluidic materials, and for engineering flexible device by utilizing the enhanced electrooptic properties.

***Acknowledgments***

We thank Dr. Yuwei Shu for editing English of our manuscript.

*Figures*

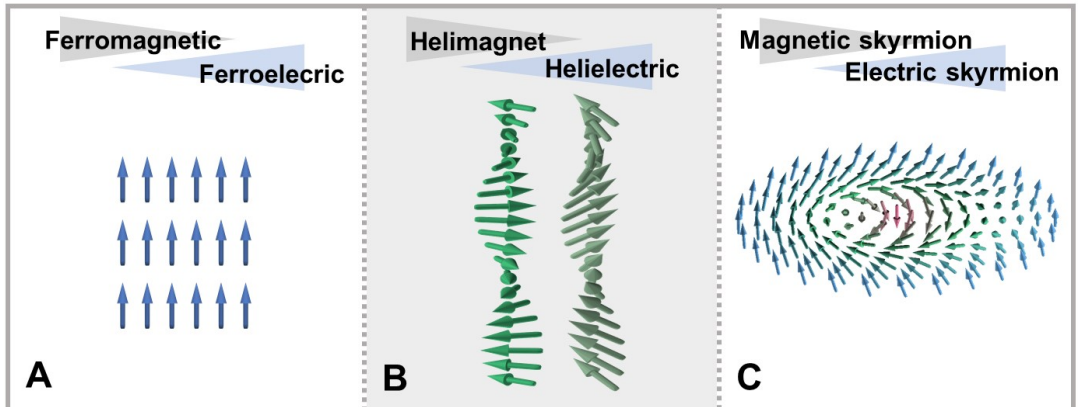

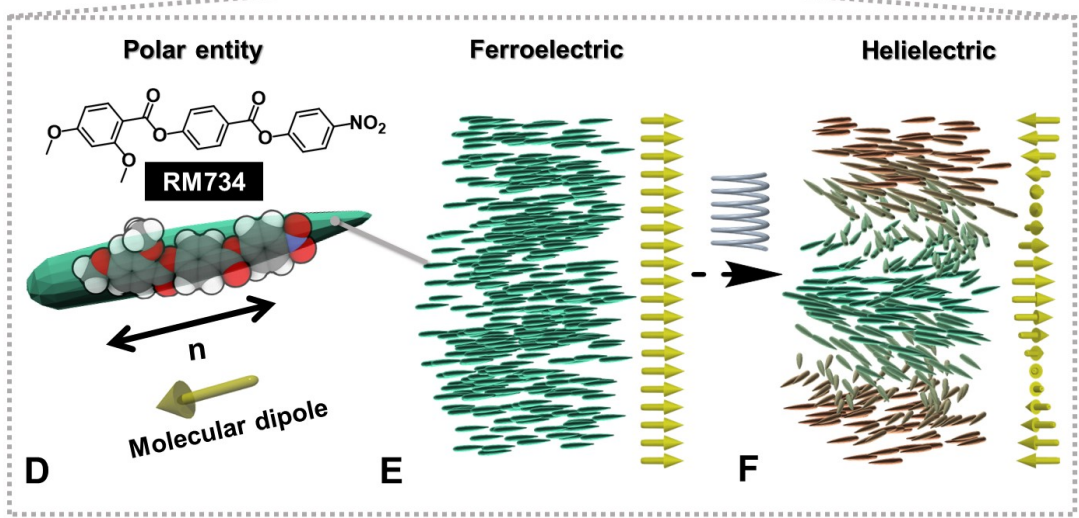

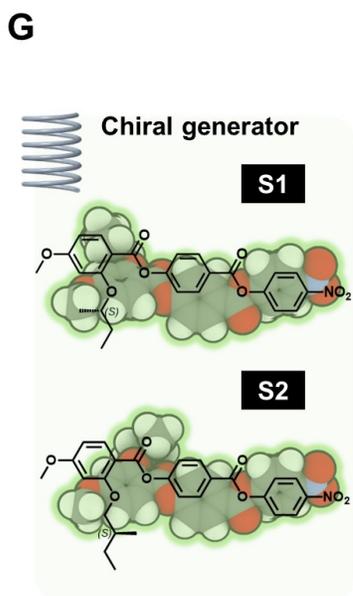

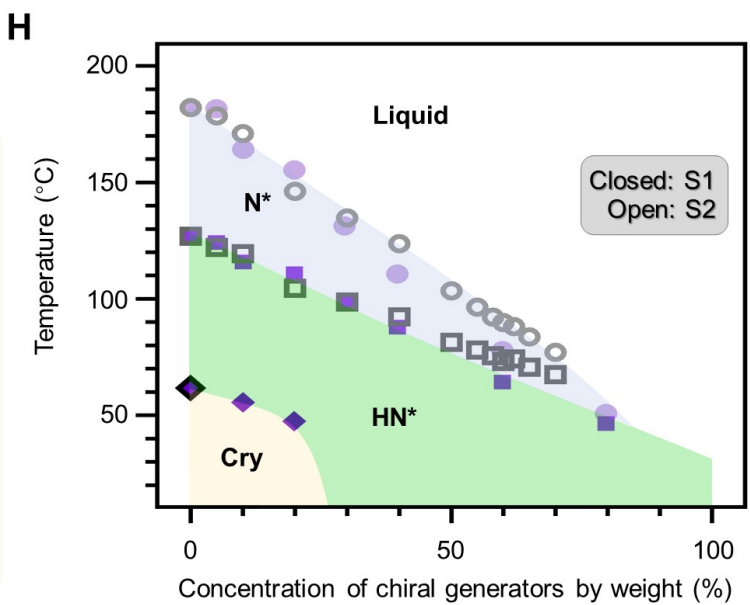

*Figure 1.* Topological analogy: electric *vs* magnetic states. Synthetic molecular library. (**A**) Uniform magnetization or polarization. (**B**) Helimagnet or helielectric states. (**C**) Magnetic or electric skyrmions. (D) Molecular structure of the polar anisotropic entity, RM734. The molecular polar dipole is nearly parallel to the long molecular axis. (E) The ferroelectric nematic state with spontaneous polarization. (F) Helielectric nematic state realized by adding chiral generators into the polar chiral nematic state. 1D polarization fields are also depicted in (E) and (F) for clarity. (G) The molecular structures of the chiral generators S1 and S2. (H) The state diagram of the two helielectric nematic materials by mixing RM734 with S1 or S2.

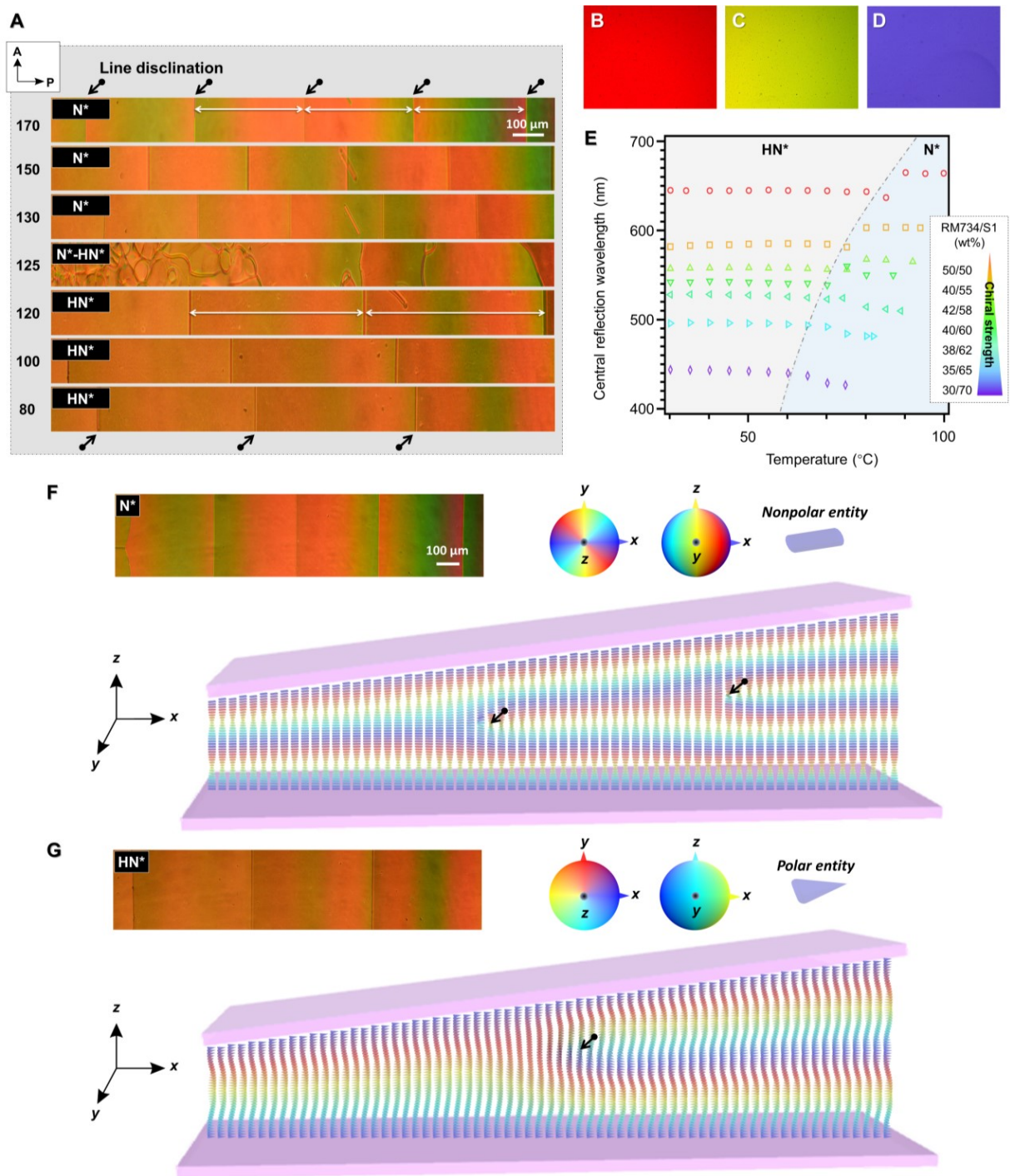

*Figure 2* General properties of helielectric nematic. (**A**) The textural evolution of RM734/S1=95/5 mixture in a wedged cell with an inclination angle of tan α = 0.0079. The numbers on left of the photographs mean the temperature in Celsius. While there is no net polarization in the N* state, helically-wound polarity field appears in the HN* state. See (**F**,**G**). The line disclinations are indicated by black arrows. In Figs. 3B-D, the POM images of the selective reflection in the HN*

state of RM734/S1=50/50 at 62°C (**B**), 42/58 at 59 °C (**C**), 30/70 at 49 °C (**D**) are shown. (**E**) The temperature dependencies of the mean visible-light reflection peaks for RM734/S1 mixtures. Interestingly, the reflection wavelength in a wide temperature range over about 80 °C barely changes (maximally within 20 nm). (**F**) POM, simulated director and polarization fields in wedged cells. The line disclinations are indicated by black arrows. (**G**) POM and simulated director field in a wedged cell in the N* state. The line disclinations are indicated by black arrows. (**F**) POM and simulated polarization field in the HN* state in the same wedged cell as in (**G**). The line disclinations are indicated by a black arrow.

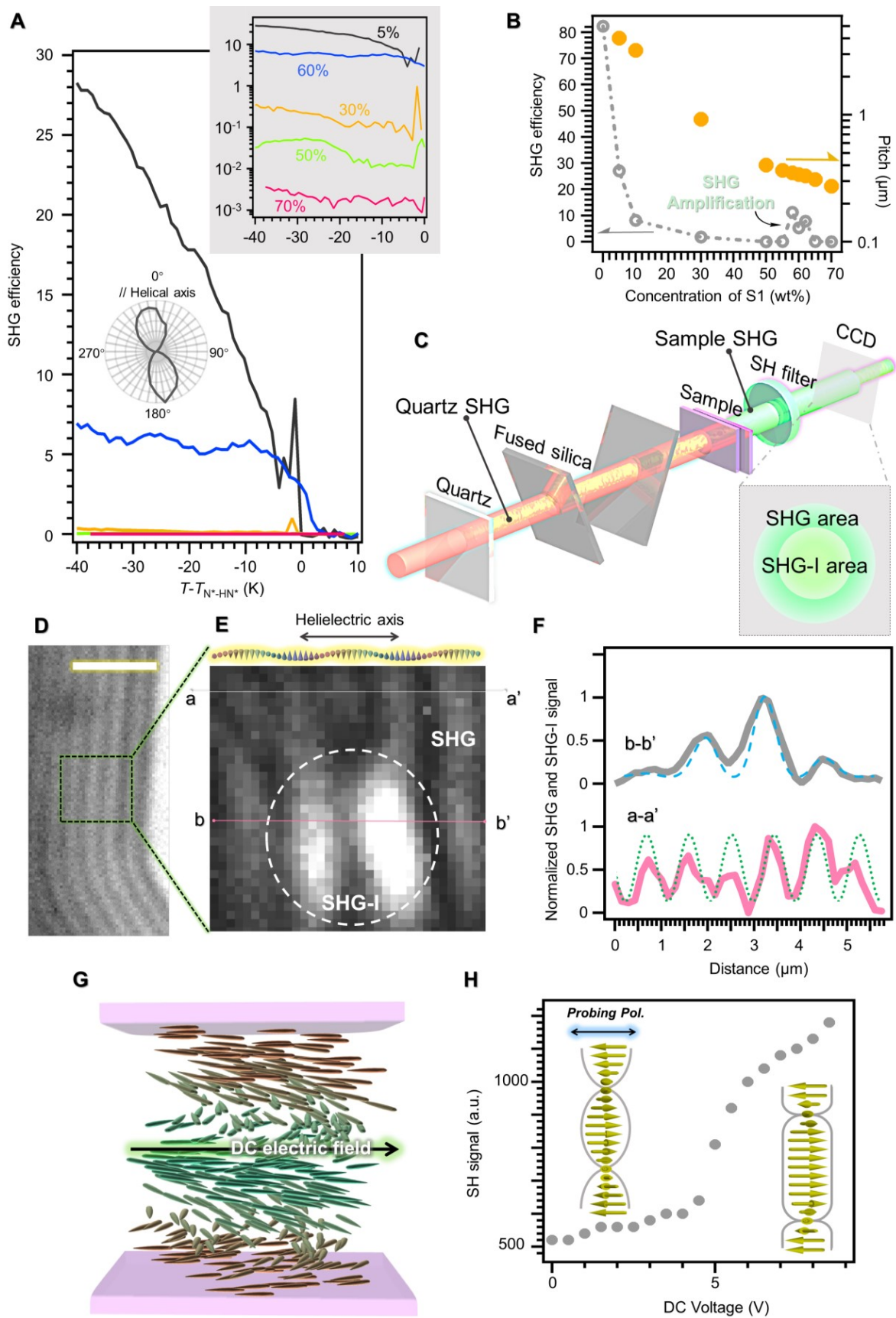

*Figure 3* SHG amplification and observation of the helielectric nematic structure. (**A**) The temperature dependencies of SH signal for for RM734/S1=95/5, 70/30, 50/50, 40/60, 30/70 mixtures. The SH signal is amplified in RM734/S1=40/60 (Also refer to (**B**)). The horizontal axis is the reduced temperature. $T_{N^*\text{-}HN^*}$ is the transition temperature between the N* and HN* states. The SHG efficiency, defined as the SH signal intensity ratio of the HN* materials to that of a Y-cut quartz plate. $T_{N^*\text{-}HN^*}$ is the transition temperature from the N* to HN* state. The inset on the right-top shows the SH signal in a log scale in the HN* state. The polar plot clearly shows the strongest SHG occurs parallel to the helical axis of the helielectric structure, suggesting the polar vector is parallel to the director. (**B**) The saturated SHG efficiency in the HN* state and helical pitch of the HN* state as a function of the weight percentage of S1. (**C**) Optical setup for the simultaneous observation of SHG and SHG-I images. In the SHG-I imaging area, the quartz and sample SH signals overlap, causing the interference of SHG. The SH signal of the polar domains interferes with that of the quartz plate constructively or destructively depending on the rotation angle of the fused silica plate. In the imaging, we choose the proper angle for best contrast of the SHG interferometry. (**D**) POM observation of the "fingerprint" helical structure. Scale bar, 5 μm. (**E**) Simultaneous observation of SHG and SHG-I microscopies. The SHG-I area, where a gaussian reference SH beam from a Y-cut quartz plate is overlapped with the sample SH signal, is indicated by a white dotted circle. The width of the image is 5.6 μm. (**F**) The SHG and SHG-I signal plots along a-a' and b-b' lines. The blue chain and green dotted line are the fitting curves to the spatial variation of the SH and SHG-I signals. The corresponding polarity field is depicted in (**E**) and (**F**). (**G**) The schematic of the HN* structure under an in-plane electric field application. (**H**) The SH signal as a function of the applied DC voltage. Over 9$V_{DC}$, electro-hydrodynamic convection occurs.

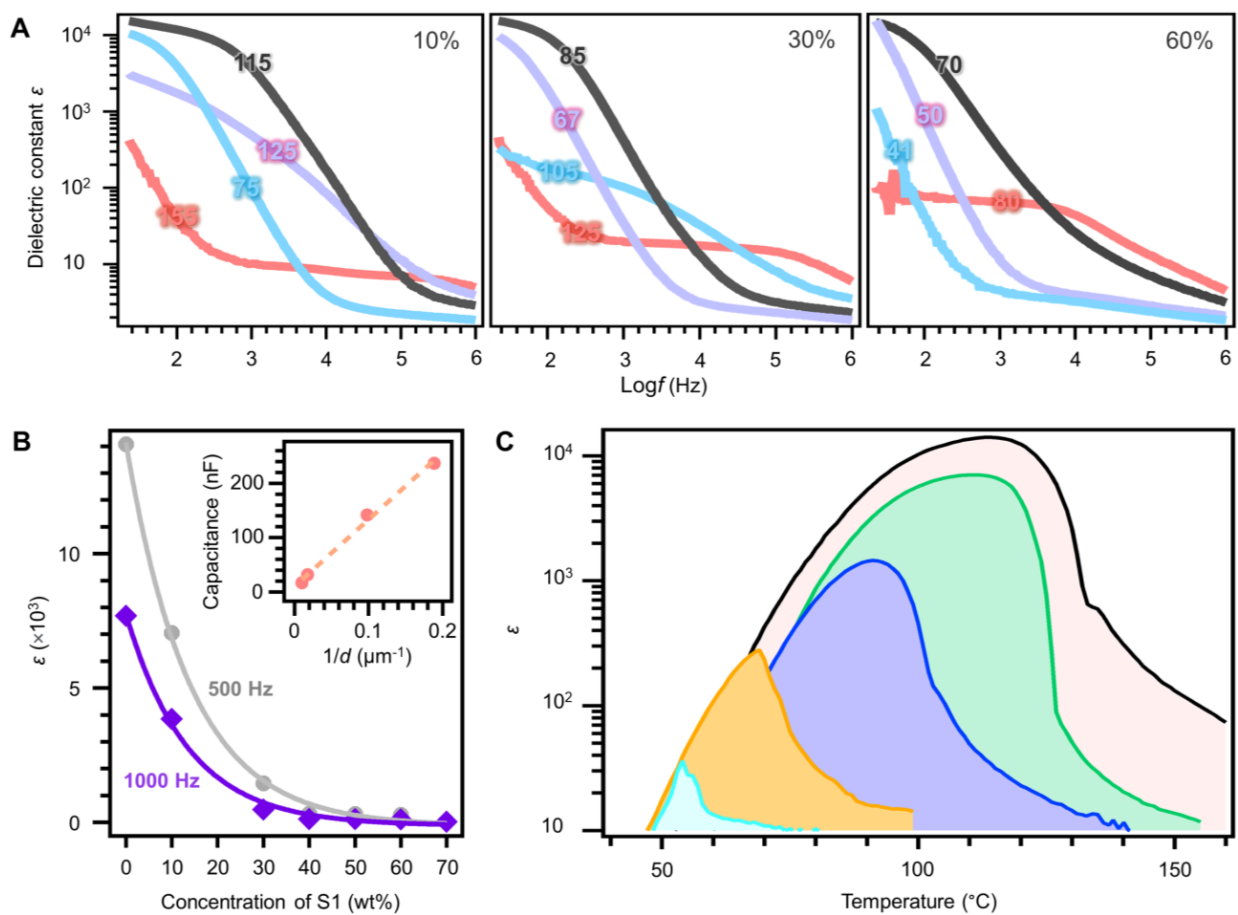

*Figure 4* Extraordinary large dielectric response. (**A**) The dielectric spectroscopy for RM734/S1=90/10, 70/30, 40/60 mixtures in their HN* state. The numbers in the figures indicate the measurement temperatures. (**B**) The maximum dielectric constant at 500 Hz and 1000 Hz as a function of the concentration of S1. The inset demonstrates the capacitance of the HN* capacitor (RM734/S1=90/10) contacting with 1cm × 1cm electrodes at different thickness *d*. The frequency of the probe electric field is 500 Hz. (**C**) The dielectric constant measured at 500 Hz as a function temperature for RM734/S1=100/0 (black), 90/10 (green), 70/30 (blue), 40/60 (orange), 30/70 (light blue) mixtures.

# Supplementary Materials for

## Observation of spontaneous helielectric nematic fluids: electric analogy to helimagnets

X. Zhao[1]†, J. Zhou[1]†, H. Nishikawa[2]†, J. Li[1], J. Kougo[1,3], Z. Wan[1], M. Huang[1,3]*, S. Aya[1,3]*

Correspondence to: huangmj25@scut.edu.cn (M.H.); satoshiaya@scut.edu.cn (S.A.)

**This PDF file includes:**

    Materials and Methods
    Figures S1-S11
    Supplementary Discussion 1-2

*Materials and Methods*

*Materials*

All commercial chemicals and solvents were used as received, unless stated otherwise. (*R*)-butan-2-ol, (*S*)-2-methylbutan-1-ol were obtained from Innochem. methyl 2-hydroxy-4-methoxybenzoate, 4-hydroxy-2-methoxybenzaldehyde, Sodium chlorite, Sodium dihydrogen phosphate were obtained from Energy Chemical. 4-nitrophenyl 4-hydroxybenzoate were obtained from Bidepharm. *N*-(3-dimethylaminopropyl)-*N'*-ethylcarbodiimide hydrochloride (EDC·HCl), and 4-dimethylaminopyridine (DMAP) were obtained from Sigma-Aldrich. Tetrahydrofuran (THF, Energy Chemical). Dichloromethane (DCM, Energy Chemical), Petroleum ether (PE, Energy Chemical) Ethyl acetate (EA, Energy Chemical), Methanol (MeOH, Energy Chemical, reagent grade), *N,N*-Dimethylformamide (DMF, Energy Chemical, reagent grade), Potassium carbonate, Potassium hydroxide were obtained from Sigma-Aldrich.

Characterizations of chemicals

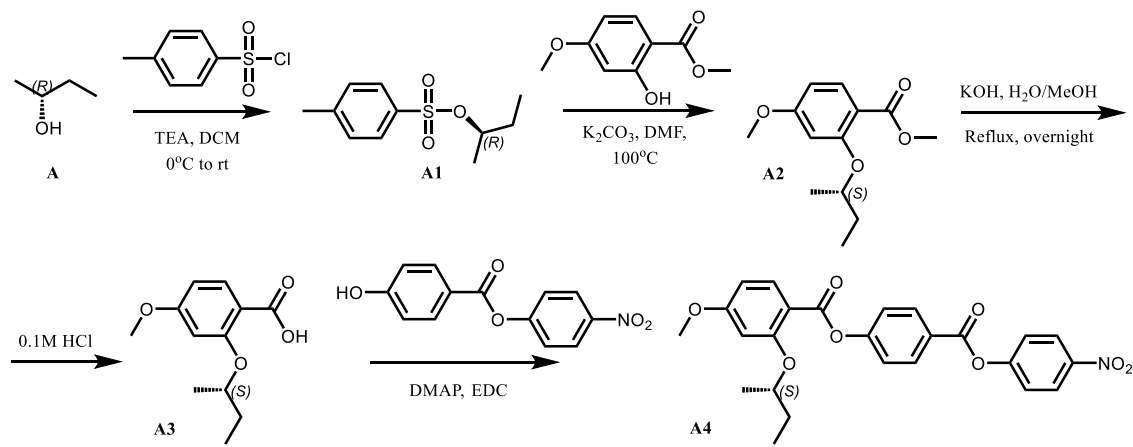

**(*R*)-sec-butyl 4-methylbenzenesulfonate (A1):** A round bottom flask was charged with (*R*)-butan-2-ol (1.00 g, 13.49 mmol), triethylamine (2.82 mL, 20.2 mmol), 50 mL DCM. 4-Methylbenzenesulfonyl chloride (3.86 g, 20.2 mmol) in 10 mL DCM was added dropwise over 20 min at 0 °C. After the mixture was stirred at room temperature overnight, the reaction mixture was concentrated in vacuum and the residue was dissolved in ethyl acetate. The resulting solution was washed with water and brine, dried with MgSO$_4$. The concentrated residue was purified by chromatography using EA/PE as eluent. Yield: 3.01 g (97.7%); appearance: colorless oily liquid. $^1$H NMR (500 MHz, Chloroform-*d*) δ 7.85 – 7.73 (m, 2H), 7.38 – 7.29 (m, 2H), 4.56 (h, *J* = 6.3

Hz, 1H), 2.44 (s, 3H), 1.69 – 1.49 (m, 2H), 1.25 (d, *J* = 6.3 Hz, 3H), 0.81 (t, *J* = 7.5 Hz, 3H). $^{13}$C NMR (126 MHz, Chloroform-*d*) δ 144.42, 134.59, 129.73, 127.68, 81.82, 29.47, 21.62, 20.29, 9.29.

**methyl (*S*)-2-(sec-butoxy)-4-methoxybenzoate (A2):** A round bottom flask was charged with **A** (1.00 g, 4.38 mmol), methyl 2-hydroxy-4-methoxybenzoate (0.960 g, 5.26 mmol), K$_2$CO$_3$ (1.82 g, 13.1 mmol), and 20 mL DMF. The solution was degassed by bubbling nitrogen for 5 min, then refluxed until the reaction was judged complete by TLC (6 - 48 h). After Cooled to room temperature, water (50 mL) was added to the residue, and the desired compound was extracted with DCM (3 × 50 mL). The organic phase was dried by anhydrous MgSO$_4$, and the solvent-removed residue was purified by chromatography. Yield: 0.82 g (78%); appearance: colorless oily liquid. $^1$H NMR (400 MHz, Chloroform-*d*) δ 7.86 – 7.77 (m, 1H), 6.50 – 6.44 (m, 2H), 4.33 (h, *J* = 6.0 Hz, 1H), 3.83 (d, *J* = 4.0 Hz, 6H), 1.80 (dqd, *J* = 13.4, 7.5, 6.0 Hz, 1H), 1.74 – 1.64 (m, 1H), 1.33 (d, *J* = 6.1 Hz, 3H), 1.00 (t, *J* = 7.4 Hz, 3H). $^{13}$C NMR (126 MHz, Chloroform-*d*) δ 166.48, 163.90, 160.03, 133.76, 113.78, 104.73, 101.60, 76.58, 55.41, 51.51, 29.20, 19.04, 9.61.

**(*S*)-2-(sec-butoxy)-4-methoxybenzoic acid (A3):** A round bottom flask was charged with **A2** (0.820 g, 3.44 mmol), KOH (2.35 g, 41.97mmol), 50 mL H$_2$O/ 50 mL MeOH. The mixture was heated for reflux overnight. After cooled to room temperature, the mixture was diluted with water (100 mL) and 1N HCl solution was added to adjust the pH to 4. The solution was extracted with ethyl acetate (3 × 50 mL). The organic phase was dried by anhydrous MgSO$_4$, and the solvent-removed residue was purified by chromatography using EA/PE as eluent. Yield: 0.76 g (98%); appearance: colorless oily liquid. 1H NMR (500 MHz, Chloroform-*d*) δ 10.17 (s, 1H), 8.14 (d, *J* = 8.8 Hz, 1H), 6.64 (dd, *J* = 8.8, 2.3 Hz, 1H), 6.52 (d, *J* = 2.3 Hz, 1H), 4.61 (h, *J* = 6.1 Hz, 1H), 3.87 (s, 3H), 1.94 – 1.74 (m, 2H), 1.43 (d, *J* = 6.2 Hz, 3H), 1.04 (t, *J* = 7.5 Hz, 3H). $^{13}$C NMR (126 MHz, Chloroform-*d*) δ 165.55, 165.00, 158.10, 135.39, 111.25, 106.87, 100.62, 78.83, 55.76, 28.98, 19.15, 9.56.

**4-((4-nitrophenoxy)carbonyl)phenyl (*S*)-2-(sec-butoxy)-4-methoxybenzoate (A4):** A round bottom flask was charged with **A3** (0.760 g, 3.39 mmol), EDC (0.971 g, 5.08 mmol), N, N-dimethylaminopyridine (0.707 g, 5.08 mmol), and 100 mL DCM was added and cooled to 0°C under nitrogen. Then 4-nitrophenyl 4-hydroxybenzoate (1.05 g, 4.07 mmol) in 20 mL DCM was added. The mixture was warmed to room temperature and stirred for 17 h. The solution was stripped of solvent by rotary evaporation, and dry-loaded onto a silica gel column for purification using EA/PE as eluent. Yield: 1.22 g (77.1%); appearance: colorless crystal. 1H NMR (400 MHz, Chloroform-*d*) δ 8.34 (d, *J* = 9.1 Hz, 2H), 8.26 (d, *J* = 8.7 Hz, 2H), 8.04 (d, *J* = 8.7 Hz, 1H), 7.41 (dd, *J* = 17.3, 8.9 Hz, 4H), 6.62 – 6.49 (m, 2H), 4.42 (q, *J* = 6.0 Hz, 1H), 3.89 (s, 3H), 1.82 (dt, *J* = 13.7, 6.8 Hz, 1H), 1.76 – 1.66 (m, 1H), 1.37 (d, *J* = 6.1 Hz, 3H), 1.01 (t, *J* = 7.4 Hz, 3H). $^{13}$C NMR (126 MHz, Chloroform-*d*) δ 165.04, 163.64, 163.30, 161.08, 156.15, 155.71, 145.43, 134.65, 131.94, 125.48, 125.30, 122.67, 122.48, 111.66, 104.96, 101.13, 76.51, 55.58, 29.20, 19.09, 9.69.

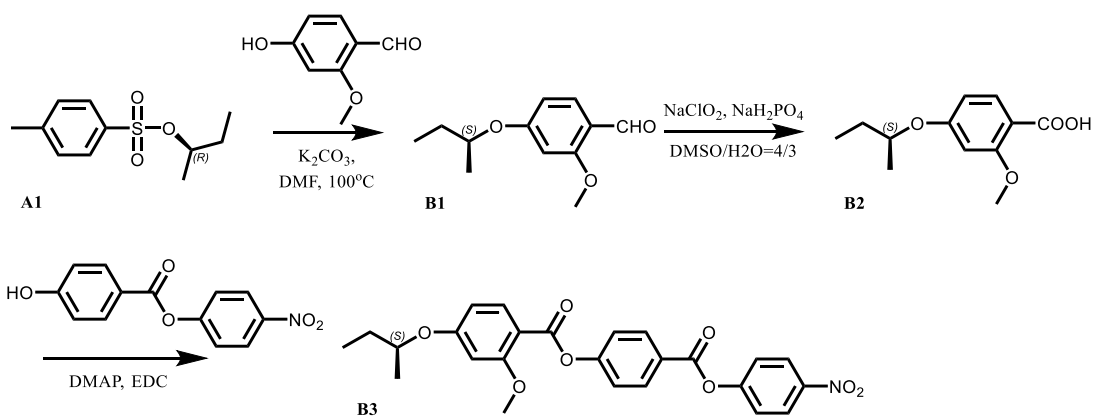

**(*S*)-4-(sec-butoxy)-2-methoxybenzaldehyde (B1):** synthesized by similar procedure as in **A2**. Yield: 78.6%; appearance: colorless oily liquid. $^1$H NMR (500 MHz, Chloroform-*d*) δ 10.27 (d, *J* = 0.8 Hz, 1H), 7.78 (d, *J* = 8.7 Hz, 1H), 6.51 (dd, *J* = 8.8, 2.3 Hz, 1H), 6.43 (d, *J* = 2.2 Hz, 1H), 4.41 (h, *J* = 6.1 Hz, 1H), 3.89 (s, 3H), 1.83 – 1.71 (m, 1H), 1.66 (dqd, *J* = 14.9, 7.4, 5.8 Hz, 1H), 1.33 (d, *J* = 6.1 Hz, 3H), 0.98 (t, *J* = 7.5 Hz, 3H). $^{13}$C NMR (126 MHz, Chloroform-*d*) δ 188.29, 165.11, 163.74, 130.72, 118.67, 106.79, 99.39, 75.45, 55.56, 29.13, 19.20, 9.71.

**(*S*)-4-(sec-butoxy)-2-methoxybenzoic acid (B2):** In a round bottom flask, sodium dihydrogen phosphate (2.30 g, 19.2 mmol) and sodium chlorite (1.52 g, 16.8 mmol) were added to a stirred solution of **B1** (1.00 g, 4.80 mmol) in a mixture of DMSO/H2O (v/v = 4/3) at 0 °C. The mixture was warmed to room temperature and stirred for 6 hours. The mixture was then diluted with water and solid NaHCO3 was added to adjust the pH to 8. The solution was washed with ethyl acetate, then the pH was adjusted to 4 by the addition of 1M HCl solution and extracted with ethyl acetate. The combined organic phase was washed with brine, dried (NaSO4) and evaporated under reduced pressure to give the compound. Yield: 1.02 g (95%); appearance: colorless oily liquid. $^1$H NMR (500 MHz, Chloroform-*d*) δ 10.27 (s, 1H), 7.78 (d, *J* = 8.7 Hz, 1H), 6.54 – 6.49 (m, 1H), 6.43 (d, *J* = 2.1 Hz, 1H), 4.41 (h, *J* = 6.1 Hz, 1H), 3.89 (s, 3H), 1.82 – 1.71 (m, 1H), 1.71 – 1.61 (m, 1H),

1.33 (d, *J* = 6.1 Hz, 3H), 0.99 (t, *J* = 7.5 Hz, 3H). ¹³C NMR (126 MHz, Chloroform-*d*) δ 165.26, 164.01, 159.57, 135.55, 109.84, 107.64, 100.07, 75.64, 56.57, 29.08, 19.12, 9.69.

**4-((4-nitrophenoxy)carbonyl)phenyl (*S*)-4-(sec-butoxy)-2-methoxybenzoate (B3):** synthesized by similar procedure as in **A4**. Yield: 79%; appearance: colorless crystal. ¹H NMR (400 MHz, Chloroform-*d*) δ 8.37 – 8.30 (m, 2H), 8.28 – 8.22 (m, 2H), 8.08 (d, *J* = 8.6 Hz, 1H), 7.48 – 7.35 (m, 4H), 6.62 – 6.50 (m, 2H), 4.44 (h, *J* = 6.1 Hz, 1H), 3.93 (s, 3H), 1.86 – 1.74 (m, 1H), 1.74 – 1.64 (m, 1H), 1.36 (d, *J* = 6.1 Hz, 3H), 1.01 (t, *J* = 7.5 Hz, 3H). ¹³C NMR (126 MHz, Chloroform-*d*) δ 164.27, 163.65, 162.79, 162.67, 156.05, 155.71, 145.43, 134.61, 131.88, 125.51, 125.30, 122.67, 122.53, 109.79, 106.11, 100.50, 75.46, 56.02, 29.12, 19.18, 9.72.

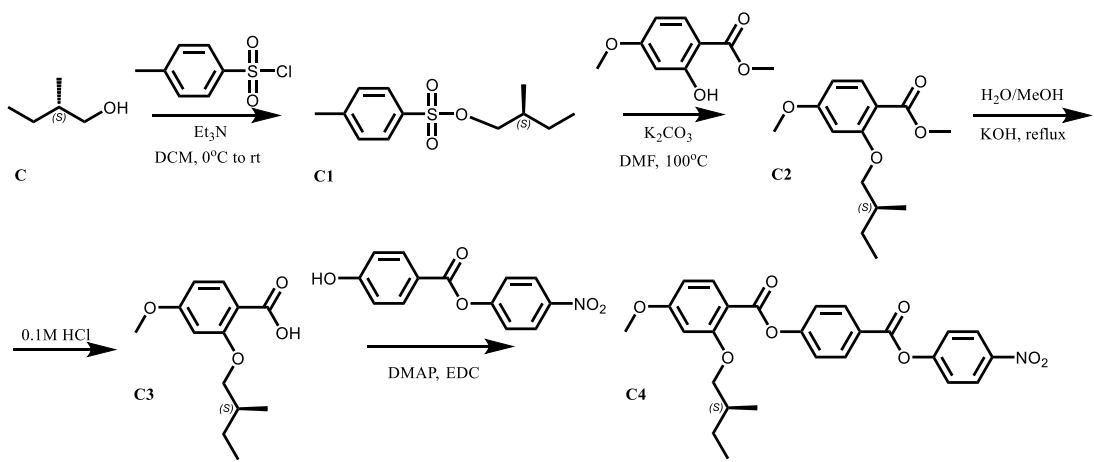

**(*S*)-2-methylbutyl 4-methylbenzenesulfonate (C1):** synthesized by similar procedure as in **A1**. Yield: 98%; appearance: colorless oily liquid. ¹H NMR (500 MHz, Chloroform-*d*) δ 7.75 (d, *J* = 8.3 Hz, 2H), 7.31 (d, *J* = 8.0 Hz, 2H), 3.89 – 3.74 (m, 2H), 2.41 (s, 3H), 1.72 – 1.61 (m, 1H), 1.41 – 1.29 (m, 1H), 1.11 (dt, *J* = 13.8, 7.5 Hz, 1H), 0.85 (s, 3H), 0.79 (d, *J* = 15.0 Hz, 3H). ¹³C NMR (126 MHz, Chloroform-*d*) δ 144.70, 133.11, 129.84, 127.86, 74.85, 34.34, 25.41, 21.61, 15.95, 10.95.

**methyl (*S*)-4-methoxy-2-(2-methylbutoxy)benzoate (C2):** synthesized by similar procedure as in **A2**. Yield: 80.2%; appearance: colorless oily liquid. ¹H NMR (500 MHz, Chloroform-*d*) δ 7.84 (d, *J* = 8.6 Hz, 1H), 6.49 – 6.42 (m, 2H), 3.88 – 3.85 (m, 1H), 3.84 (d, *J* = 7.8 Hz, 6H), 3.77 (dd, *J* = 8.7, 6.6 Hz, 1H), 1.92 (dq, *J* = 13.1, 6.4 Hz, 1H), 1.64 – 1.56 (m, 1H), 1.32 (dd, *J* = 13.7, 7.4 Hz, 1H), 1.06 (d, *J* = 6.7 Hz, 3H), 0.95 (t, *J* = 7.5 Hz, 3H). ¹³C NMR (126 MHz, Chloroform-*d*) δ

166.52, 164.13, 160.93, 133.85, 112.56, 104.49, 99.64, 73.50, 55.45, 51.57, 34.77, 26.05, 16.56, 11.36.

**(*S*)-4-methoxy-2-(2-methylbutoxy)benzoic acid (C3):** synthesized by similar procedure as in **A3**. Yield: 96.8%; appearance: colorless oily liquid. $^1$H NMR (500 MHz, Chloroform-*d*) δ 7.82 (d, *J* = 8.5 Hz, 1H), 6.51 – 6.36 (m, 2H), 3.87 – 3.73 (m, 8H), 1.90 (dddd, *J* = 12.2, 7.8, 6.6, 5.6 Hz, 1H), 1.59 (dqd, *J* = 13.1, 7.5, 5.5 Hz, 1H), 1.29 (dp, *J* = 13.5, 7.5 Hz, 1H), 1.04 (d, *J* = 6.9 Hz, 3H), 0.93 (t, *J* = 7.5 Hz, 3H). $^{13}$C NMR (126 MHz, Chloroform-*d*) δ 165.37, 165.05, 159.13, 135.48, 110.39, 106.58, 99.35, 74.83, 55.75, 34.42, 26.08, 16.59, 11.21.

**4-((4-nitrophenoxy)carbonyl)phenyl (*S*)-4-methoxy-2-(2-methylbutoxy)benzoate (C4):** synthesized by similar procedure as in **A4**. Yield: 77.5%; appearance: colorless crystal. $^1$H NMR (400 MHz, Chloroform-*d*) δ 8.37 – 8.30 (m, 2H), 8.30 – 8.23 (m, 2H), 8.06 (d, *J* = 8.7 Hz, 1H), 7.47 – 7.35 (m, 4H), 6.56 (dd, *J* = 8.8, 2.3 Hz, 1H), 6.52 (d, *J* = 2.3 Hz, 1H), 3.96 – 3.82 (m, 5H), 1.99 – 1.88 (m, 1H), 1.67 – 1.59 (m, 1H), 1.36 – 1.28 (m, 1H), 1.06 (d, *J* = 6.8 Hz, 3H), 0.93 (t, *J* = 7.5 Hz, 3H). $^{13}$C NMR (126 MHz, Chloroform-*d*) δ 165.26, 163.63, 163.35, 161.95, 156.13, 155.71, 145.43, 134.73, 131.97, 125.52, 125.29, 122.66, 122.48, 110.59, 104.92, 99.50, 73.53, 55.61, 34.76, 26.00, 16.60, 11.34.

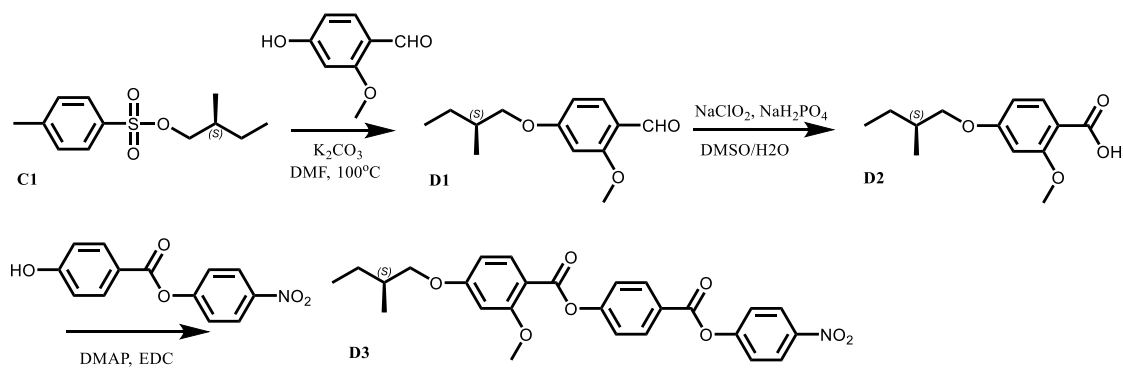

**(*S*)-2-methoxy-4-(2-methylbutoxy)benzaldehyde (D1):** synthesized by similar procedure as in **B1**. Yield: 79.8%; appearance: colorless oily liquid. $^1$H NMR (500 MHz, Chloroform-*d*) δ 10.27 (d, *J* = 0.8 Hz, 1H), 7.78 (d, *J* = 8.6 Hz, 1H), 6.52 (ddd, *J* = 8.7, 2.2, 0.8 Hz, 1H), 6.44 (d, *J* = 2.2 Hz, 1H), 3.89 (s, 3H), 3.88 – 3.85 (m, 1H), 3.80 (dd, *J* = 9.0, 6.5 Hz, 1H), 1.93 – 1.82 (m, 1H), 1.57 (dqd, *J* = 13.1, 7.5, 5.5 Hz, 1H), 1.27 (dq, *J* = 13.5, 7.5 Hz, 1H), 1.03 (d, *J* = 6.8 Hz, 3H), 0.96 (t, *J* = 7.5 Hz, 3H). $^{13}$C NMR (126 MHz, Chloroform-*d*) δ 188.26, 166.04, 163.67, 130.61, 118.81, 106.36, 98.25, 73.20, 55.58, 34.65, 26.05, 16.47, 11.30.

**(S)-2-methoxy-4-(2-methylbutoxy)benzoic acid (D2):** synthesized by similar procedure as in **B2**. Yield: 96.5%; appearance: colorless oily liquid. $^1$H NMR (500 MHz, Chloroform-$d$) δ 10.63 (s, 1H), 8.10 (d, $J$ = 8.8 Hz, 1H), 6.62 (d, $J$ = 11.0 Hz, 1H), 6.54 (d, $J$ = 2.2 Hz, 1H), 4.04 (s, 3H), 3.91 – 3.85 (m, 1H), 3.85 – 3.77 (m, 1H), 1.94 – 1.83 (m, 1H), 1.62 – 1.52 (m, 1H), 1.33 – 1.23 (m, 1H), 1.03 (d, $J$ = 6.8 Hz, 3H), 0.96 (t, $J$ = 7.5 Hz, 3H). $^{13}$C NMR (126 MHz, Chloroform-$d$) δ 165.59, 164.95, 159.61, 135.46, 110.04, 107.11, 98.97, 73.34, 56.61, 34.62, 26.03, 16.46, 11.29.

**4-((4-nitrophenoxy)carbonyl)phenyl (S)-2-methoxy-4-(2-methylbutoxy)benzoate (D3):** synthesized by similar procedure as in **B3**. Yield: 78%; appearance: colorless crystal. $^1$H NMR (500 MHz, Chloroform-$d$) δ 8.36 – 8.31 (m, 2H), 8.28 – 8.23 (m, 2H), 8.08 (d, $J$ = 8.7 Hz, 1H), 7.46 – 7.37 (m, 4H), 6.60 – 6.52 (m, 2H), 3.95 (s, 3H), 3.91 (dd, $J$ = 9.0, 6.0 Hz, 1H), 3.83 (dd, $J$ = 9.0, 6.5 Hz, 1H), 1.91 (dq, $J$ = 13.1, 6.6 Hz, 1H), 1.64 – 1.57 (m, 1H), 1.36 – 1.29 (m, 1H), 1.05 (d, $J$ = 6.7 Hz, 3H), 0.98 (t, $J$ = 7.5 Hz, 3H). $^{13}$C NMR (126 MHz, Chloroform-$d$) δ 165.21, 163.65, 162.82, 162.58, 156.03, 155.71, 145.44, 134.60, 131.88, 125.53, 125.30, 122.67, 122.54, 110.01, 105.53, 99.40, 73.22, 56.05, 34.68, 26.08, 16.51, 11.33.

*Sample Preparation*

The polar helical materials were realised by mixing a ferroelectric nematic LC (RM734 in Fig. 2A) with the commercial and four home-made chiral generators (S811, E1, E2, S1 and S2, Figs. 2A and S2) independently by weight. S1 and S2 were designed to have the same backbone of mesogen with $N_F$ LC. The values of the optical rotation for S1 and S2 are 0.006°. The mixtures were introduced to home-made LC slabs with controlled thickness and alignment conditions.

*SHG interference (SHG-I) measurement*

An SHG-I measurement system is built as discussed in Figs. 4A and S10, Supplementary Discussions 2. The fundamental beam is emitted from a Q-switched pulsed laser (MPL- 277 III-1064-20μJ) with a central wavelength of 1064 nm, pulse duration of 5 ns, and 100 Hz repetition. The SH light is detected in the transmission geometry by a photomultiplier tube (DH-PMT-D100V, Daheng Optics) or a scientific CMOS camera (Zyla-4.2P-USB3, Andor).

*Dielectric spectroscopy*

The dielectric spectroscopy was conducted by using an LCR meter (4284A, Agilent). The data collections of the frequency and temperature sweeping are automated by a home-made software written in Labview.

*Calculation of the director and polarity orientation profile in the wedge slab*

To simulate both the structures of N* and HN* states in wedge slabs, we perform the relaxation minimization of the Ginzburg–Landau functional based on a Landan-de Gennes modelling *(28,29)*. For the N* state, we wrote the free energy as expansions in the orientational order in terms of a symmetric and traceless tensor $Q_{ij} = S(n_i n_j - \delta_{ij}/3)$, where i, j are the Cartesian coordinates and $S$ the scalar order parameter. The head-to-tail invariance is guaranteed here. To the lowest relevant orders, the free energy density is written as:

$f = f_L + f_e + f_s,$

$f_L = A_0(1 - U/3)Q_{\alpha\beta}^2/2 - A_0 U Q_{\alpha\beta} Q_{\beta\zeta} Q_{\zeta\alpha}/3 + A_0 U (Q_{\alpha\beta}^2)^2/4,$

$f_e = L(\partial_\beta Q_{\alpha\beta})^2/2 + 2q_0 L \epsilon_{\alpha\zeta\delta} Q_{\alpha\beta} \partial_\zeta Q_{\delta\beta},$

$f_s = W(Q_{\alpha\beta} - Q_{\alpha\beta}^0)^2/2$

where $f_L$, $f_e$ and $f_s$ are the Landau, the elastic and surface anchoring free energies. $A_0$ and $U$ phenomenological parameters depending on temperature and pressure. $U$ controls the magnitude of the order. $L$, $\epsilon_{\alpha\zeta\delta}$, $W$ and $Q_{\alpha\beta}^0$ are the elastic constant, Levi-Civita tensor, anchoring coefficient, and surface $Q$-tensor component. We used typical material parameters for the calculation: $A_0$, $U$ and $W$ are 1, 6 and 100, respectively. $L$ is used in the range of 0.02 to 1.

For the calculation of the topological structure for the HN* state, we start with the topological structure of the N* structure obtained as described above. To include the vectorized polar nature that breaks the head-to-tail invariance, we write the free energy density as expansions of the director orientation, $n$, as follows:

$f = \frac{1}{2}K_{11}(\text{div }\mathbf{n})^2 + \frac{1}{2}K_{22}(\mathbf{n} \cdot (\text{curl }\mathbf{n}))^2 + \frac{1}{2}K_{33}(\mathbf{n} \times (\text{curl }\mathbf{n}))^2 - \frac{1}{2}K_{22}'\mathbf{n} \cdot (\nabla \times \mathbf{n}).$

Based on the result of the SHG-I measurement, the surface polarization in the adjacent domains align in the same direction (Fig. S9). Therefore, we set the surface polarity orientation, equivalent to the director vector, to be identical for all the bottom and top surfaces. For simplicity, we relax the above director-based free energy to get the equilibrium structure. Though the assumption is rough, but still the simulated orientational pattern captures the essential structural difference between the N* and HN* state.

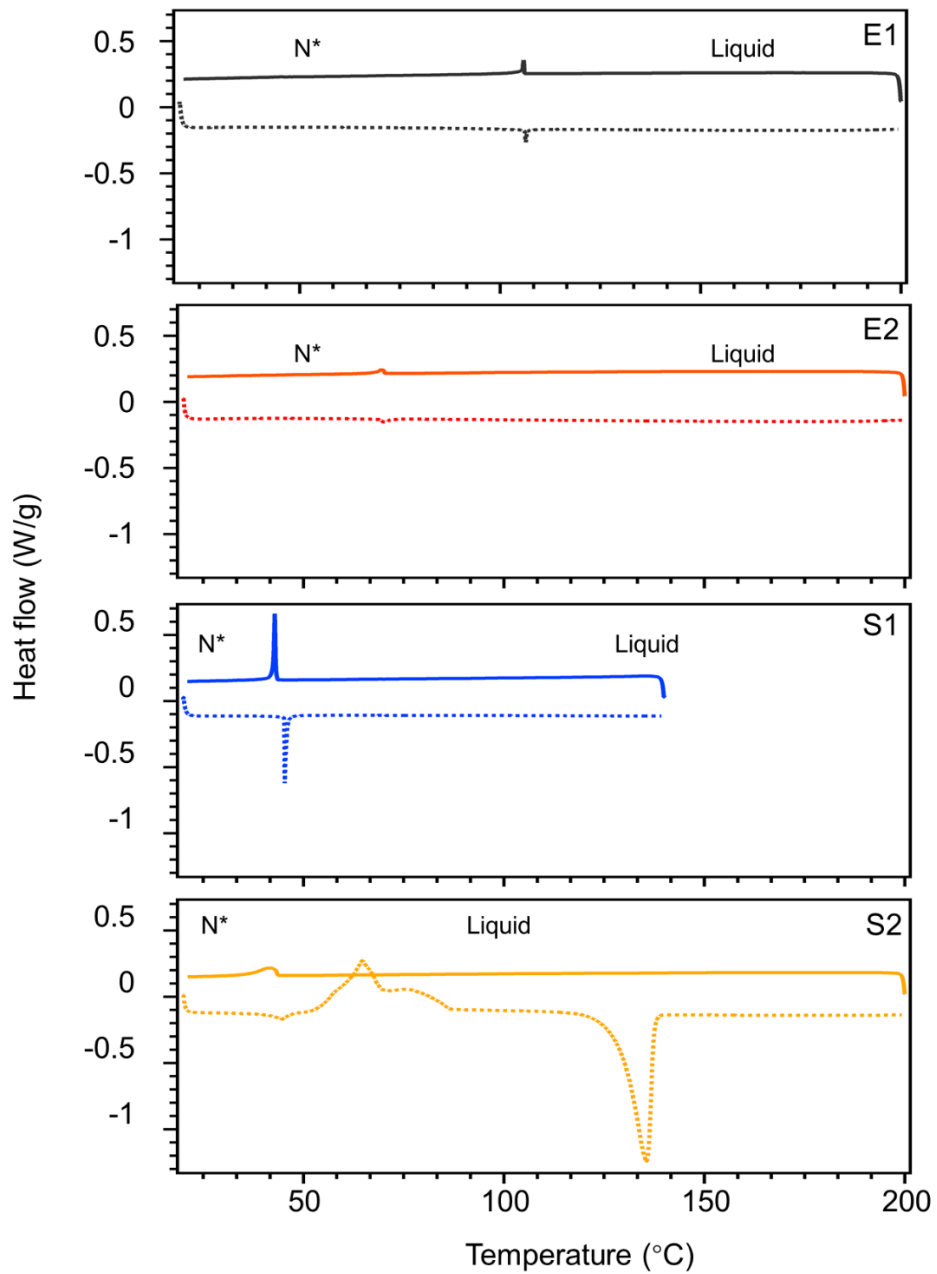

*Figure* **S1.** Phase properties and polarity of the neat chiral generators.

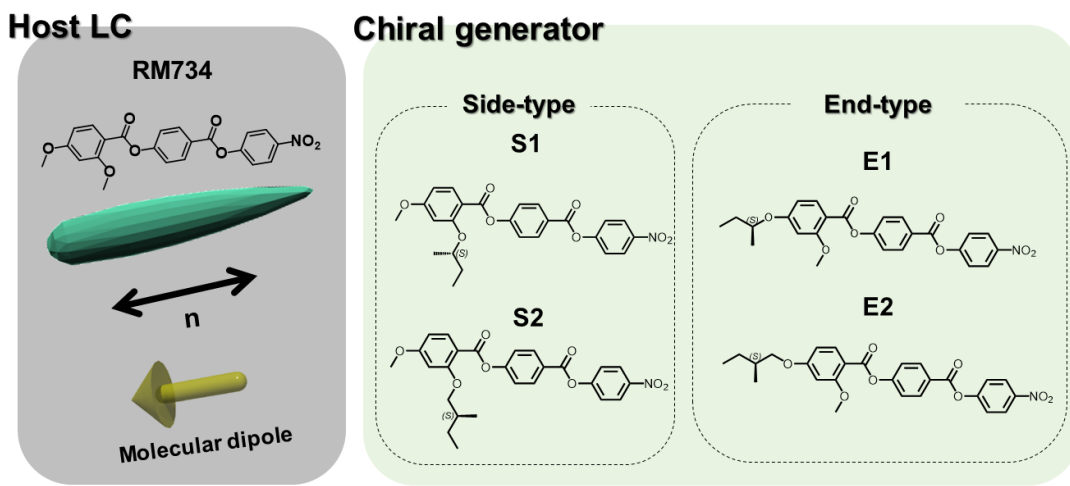
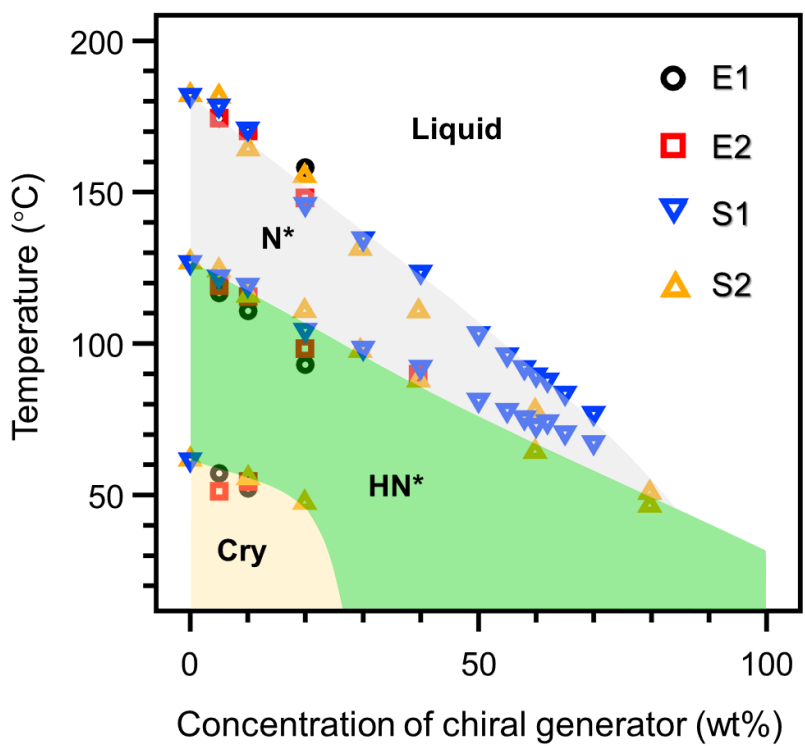

*Figure* **S2:** State diagrams of four types of mixtures. The molecular structures of the chiral generators, E1, E2, S1 and S2, are shown. Independent of the structure of the chiral generators, the phase state is generic.

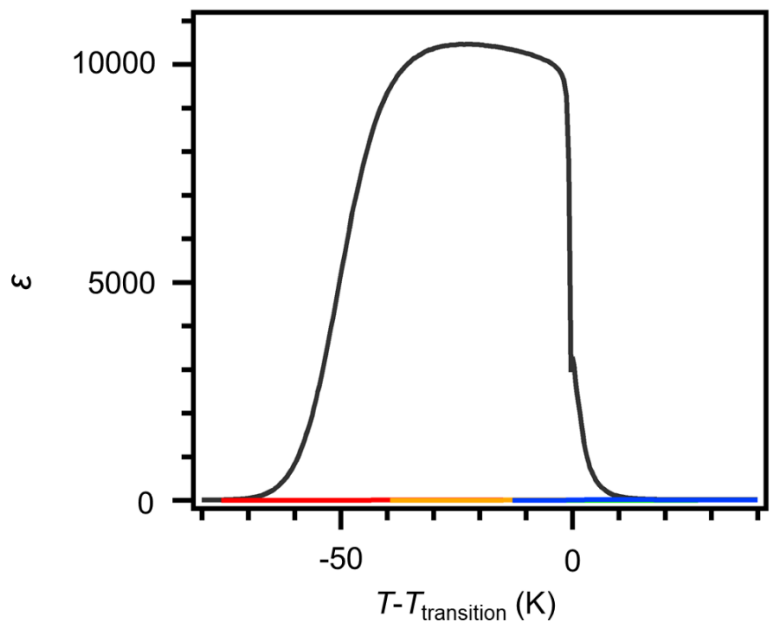

*Figure* **S3:** The dielectric constant as a function of the reduced temperature for pure E1 (red), E2 (orange), S1 (green), S2 (blue) and RM734 (black). $T_{transition}$ is the transition temperature between the N* and HN* states for RM734, and the transition temperature between the isotropic liquid and N* states for the chiral generators.

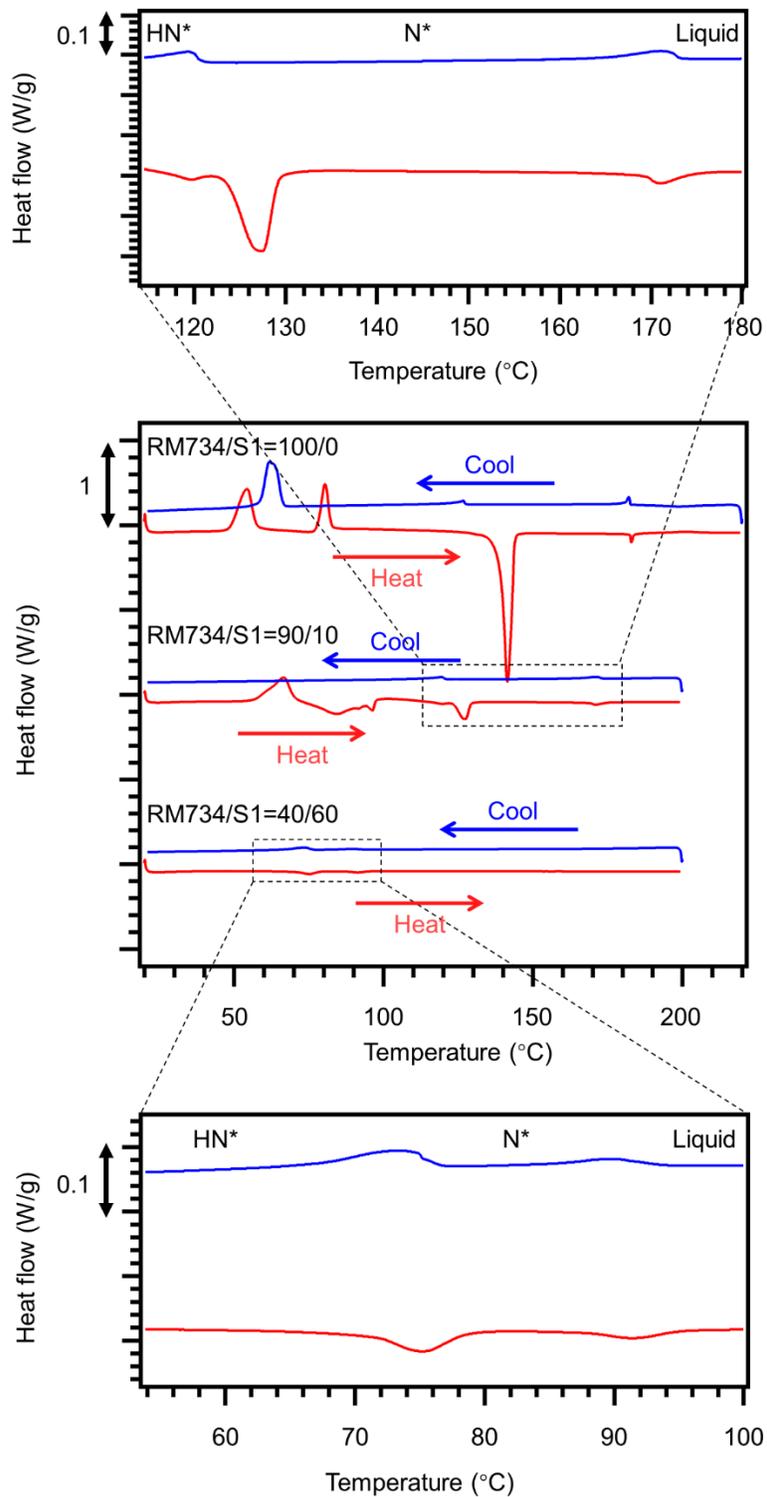

***Figure* S4:** DSC traces of mixtures of RM734/S1=100/0, 90/10, 40/60 (wt%/wt%). Clearly, increasing the amount of S1 suppresses the crystallization, thereby expanding the temperature range of the helielectric state.

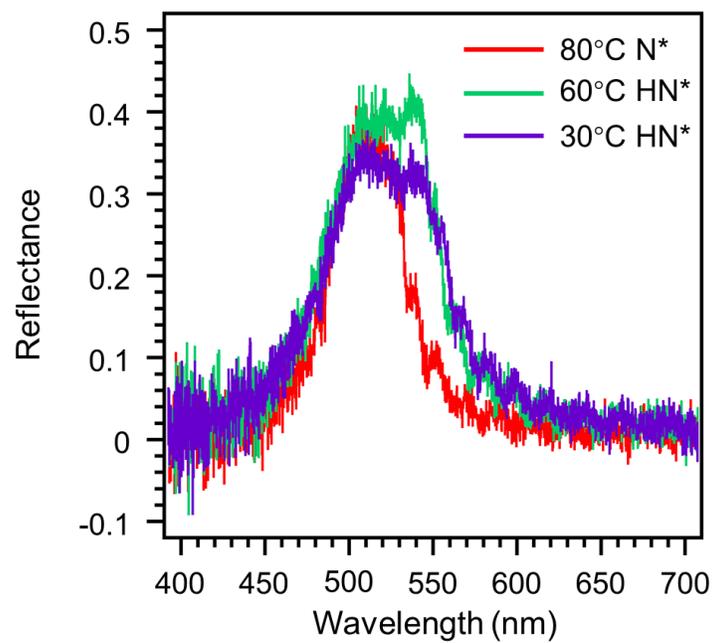

*Figure* **S5:** The reflection spectrum for RM734/S1=38/62 mixture.

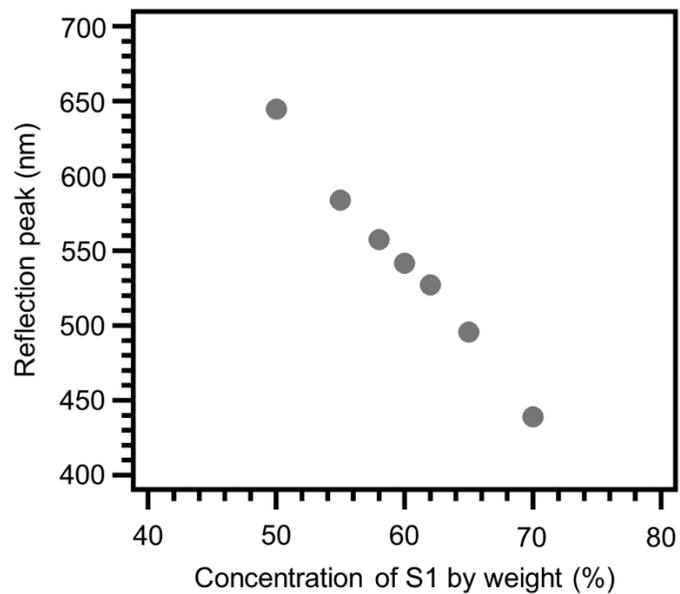

*Figure* **S6:** Mean visible-light reflection wavelength form the HN* phase for RM734/S1 mixtures as a function of the weight percentage of S1.

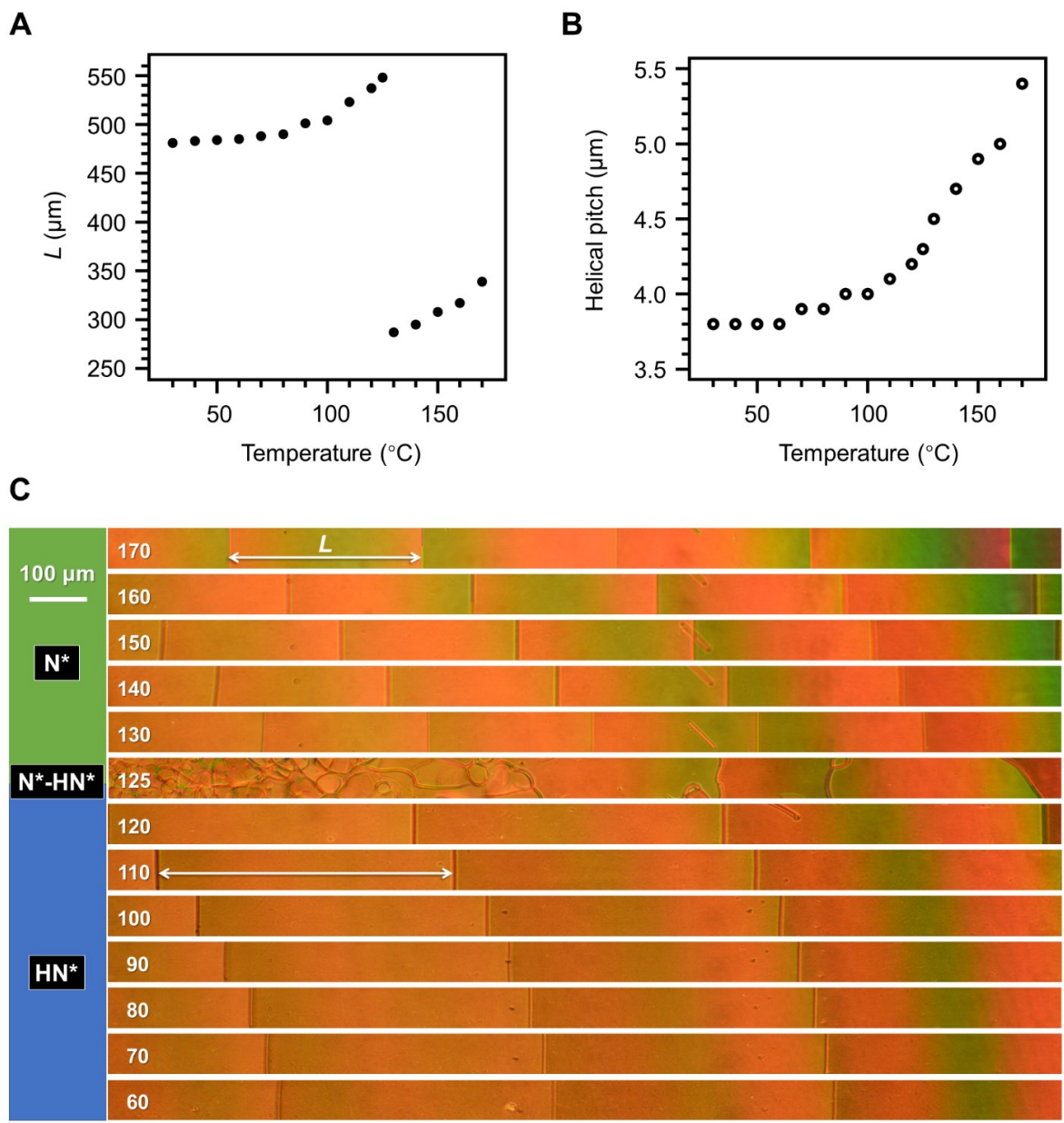

*Figure* S7: (A) The temperature dependence of the width of the Cano line in a long-pitch HN* material (RM734/S1=95/5). (B) The temperature dependence of the helical pitch. (C) The texture

evolution upon a cooling procedure from the N* to HN* phase. The numbers in the images mean the temperature.

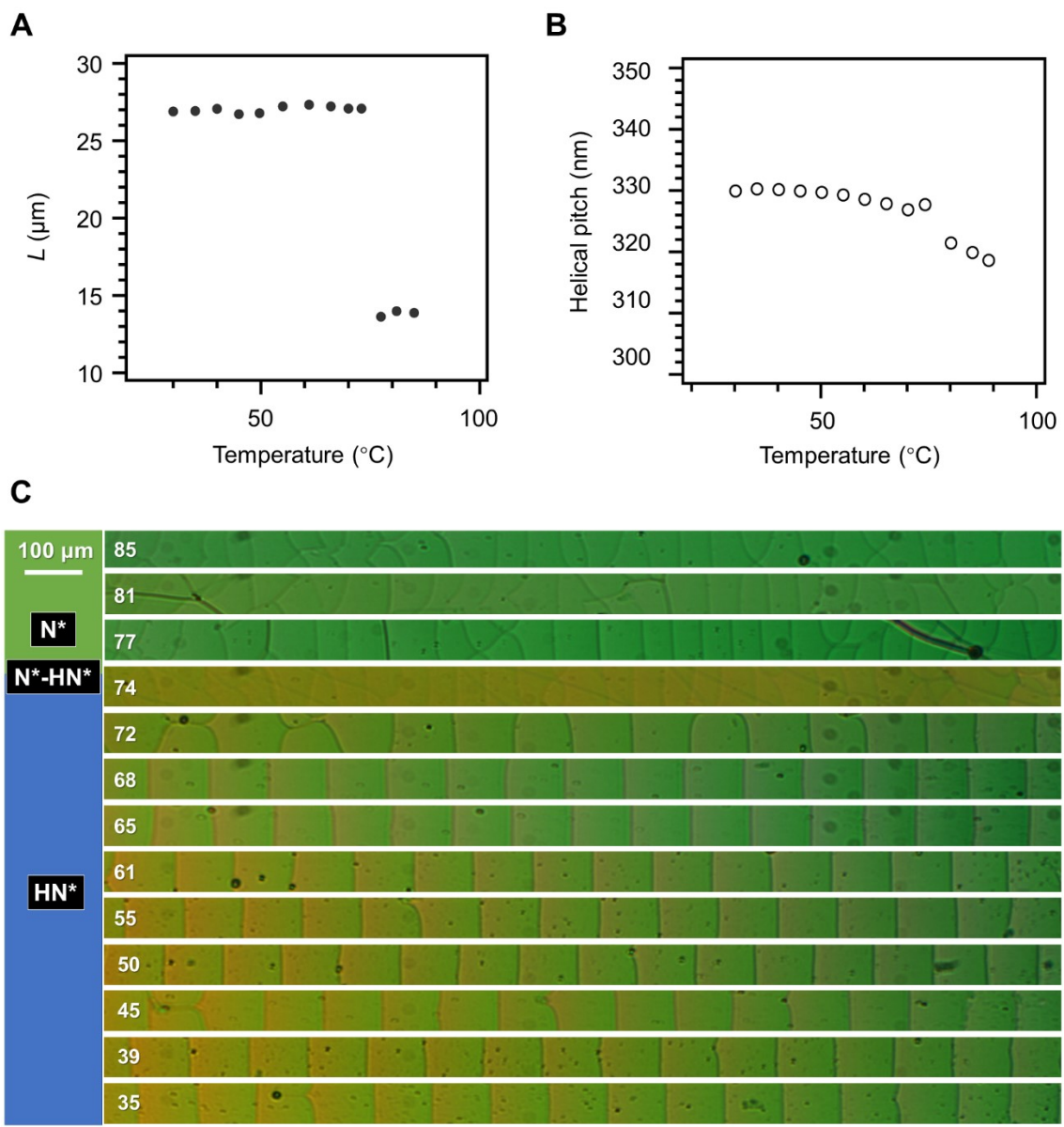

*Figure* **S8:** (A) The temperature dependence of the width of the Cano line in a short-pitch HN* material (RM734/S1=38/62). (B) The temperature dependence of the helical pitch. (C) The texture evolution upon a cooling procedure from the N* to HN* phase. The numbers in the images mean the temperature.

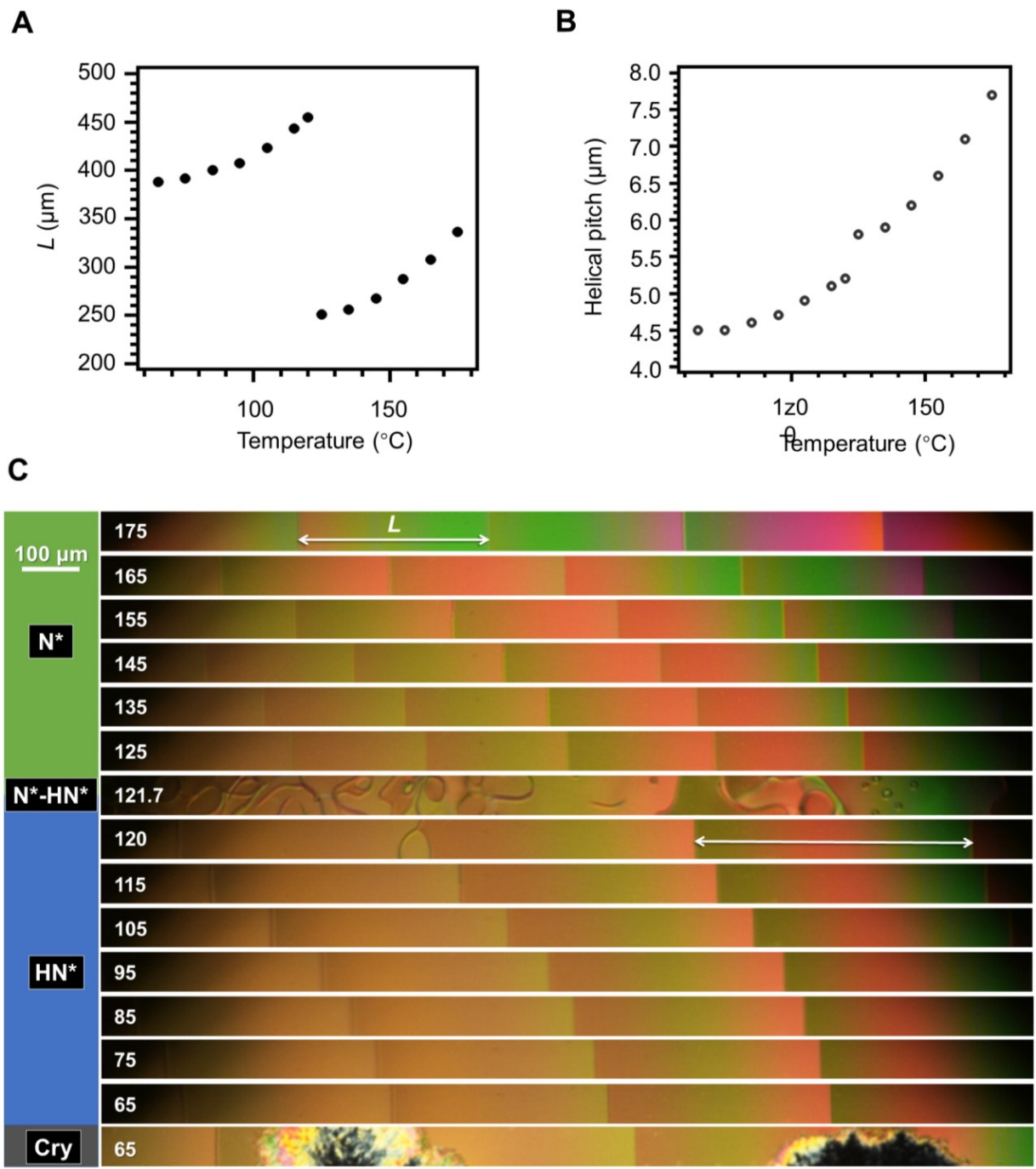

*Figure* S9: (A) The temperature dependence of the width of the Cano line in a long-pitch HN* material (RM734/S811=98.2/1.8). (B) The temperature dependence of the helical pitch. (C) The texture evolution upon a cooling procedure from the N* to HN* phase. The numbers in the images mean the temperature.

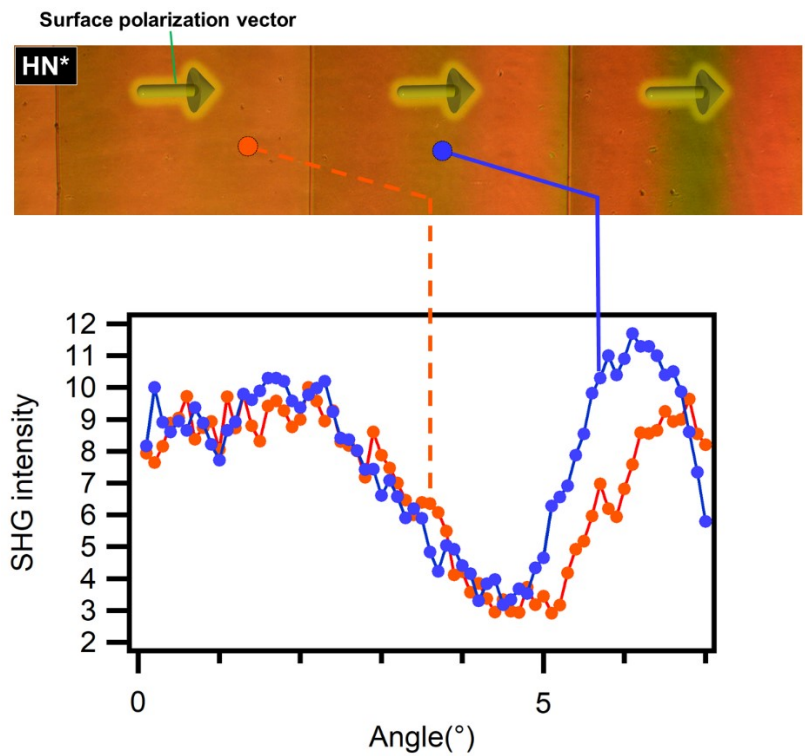

*Figure* **S10**: SHG interferometry curves for surface polarizations in adjacent domains separated by a Grandjean-Cano line.

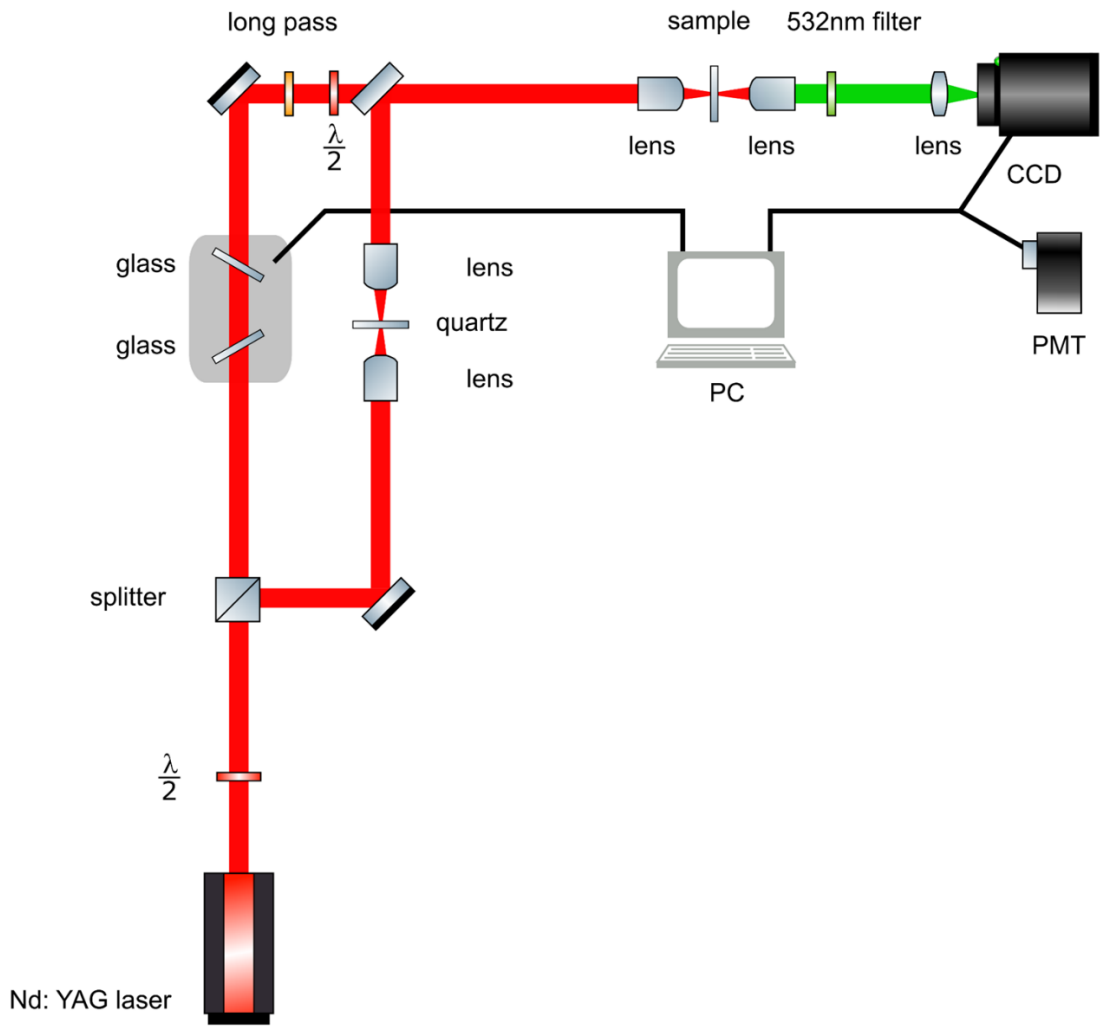

***Figure* S11:** The schematic of the interference SHG measuring system. PMT and CCD are used for point- and imaging-SHG measurements.

**Supplementary Discussion 1: The doping effect of chiral generators on the matter phase state.**

As highlighted in the main text, the mixture materials with doping distinct chiral generators have given almost the same phase state diagram (Figs. 1H and S2). Among the four chiral generators (Fig. S2), the molecular structures of S1, S2, E1 and E2 are similar to RM734, processing slight polarity as confirmed by the SH signal as strong as 0.1 of Y-cut quartz plate. We confirmed that they exhibit distinct chiral strength. These results mean that the matter phase state is determined solely by how much the volume is occupied by the chiral generators in the mixture materials. The chiral strength does not affect the thermodynamic stability of each phase. It just slightly modifies how easy to form a crystalline phase at low concentrations of the chiral generators, which might be caused by the geometrical hindrance for the crystal nucleation.

**Supplementary Discussion 2: Second harmonic generation interference measurement and microscopy**

Optical second-harmonic generation (SHG) is one of the second-order nonlinear optical phenomena, where the input fundamental beam at the wavelength $\lambda$ is converted to an SH wave at the wavelength $\lambda/2$, i.e. the optical frequency doubles. To produce SH signal, a matter interacting with light should lack the inversion center. In the present study, the systems change their point group from $D_\infty$ to $C_\infty$ upon the transition from the N* to HN* state. This symmetry variation permits the HN* state to emit SH signal as observed. To understand the spatial distribution of the polarity vector in the HN* state, we conducted SHG interference (SHG-I) measurement and microscopy. Figure S11 demonstrates the schematic of the SHG-I measurement system. We use two fused silica plates for generating the phase difference between the SH signal from the sample and the reference Y-cut quartz plate and rotate them in opposite directions to avoid the variation of the optical path during their rotation. For the SHG-I microscopy, we choose several interference conditions for the best imaging contrast by finding the corresponding rotating angles of the fused silica plates. For the simultaneous observation of both the SHG and SHG-I microscopy imaging in a single acquisition, we deliberately reduce the beam size of the SHG spot from the quartz plate. Therefore, the SHG-I imaging occurs on the CCD camera in a small area, and the other area just show the normal SHG imaging. The SHG(-I) microscopy images are recorded by Zyla-4.2P-USB3 (Andor) and subjected the subsequent analyses.